\documentclass[rnote]{aa}
\usepackage{graphicx}
\usepackage{txfonts}

\begin{document}
\title{A Detection of an Anti-correlated Hard X-ray Lag in AM Herculis}
\author{
	K. Sriram\inst{1} \and 
	C. S. Choi\inst{1} \and 
	A. R. Rao\inst{2} 
}
\institute{
International Center for Astrophysics, 
Korea Astronomy and Space Science Institute, 
Daejeon 305-348, South Korea
\and 
Tata Institute of Fundamental Research, Mumbai 400005, India
}

\date{}

\abstract
%Context
{Earlier cross-correlation studies for AM Her were performed in various energy range 
from optical to X-ray and suggested that it mostly shows a high level of
correlation but on occasion it shows a low level of correlation or uncorrelation. 
}
%Aims
{To investigate the degree of correlation between soft (2-4 keV) and hard (9-20 keV) X-rays, 
we perform the cross-correlation study of the X-ray data sets of AM Her obtained with {\it RXTE}.
}
% Methods
{We cross-correlate the background-subtracted soft and hard X-ray light curves using the XRONOS
program crosscor and fit a model to the obtained cross-correlation functions.
}
% Results
{We detect a hard X-ray lag of $192\pm33$ s in a specific section of
energy-dependent light curve, where the soft X-ray (2-4 keV) intensity decreases
but the hard X-ray (9-20 keV) intensity increases. 
From a spectral analysis, we find that the X-ray emission temperature increases 
during the anti-correlated intensity variation.
In two other observations, the cross-correlation functions
show a low level of correlation, which is consistent with the earlier
results performed in a different energy range.
}
%Conclusions
{We report a detection of an anti-correlated hard X-ray lag of $\sim$190 s from the 
proto-type polar AM Her. 
The hard X-ray lag is detected for the first time in the given energy range, and it
is the longest lag among those reported in magnetic cataclysmic variables.
We discuss the implications of our findings regarding the origin of the hard X-ray
lag and the anti-correlated intensity variation.
}

\authorrunning{Sriram, Choi, Rao}
\titlerunning{Anti-correlated Hard Lag in polar AM Her}

\maketitle

\section{Introduction}
\label{s:intro}
Magnetic cataclysmic variables (mCVs) are binaries consisting of a
magnetized white dwarf and a late spectral type companion.
The magnetic field strength is one of the important parameters when categorizing
them into subclasses. In polars (AM Her-type, B $\sim$ 10 - 230 MG) and intermediate polars 
(DQ Her-type, B $\sim$ 1 - 10 MG), accretion stream  or flow 
from the companion follows field lines which are connected to the magnetic poles of the white
dwarf (WD). 
As the magnetically channeled material accretes onto the magnetic poles (known 
as an accretion column), it experiences a strong standing shock above the magnetic
poles of the WD surface, subsequently reaching a high temperature ($\sim$10$^{8}$ K).
The hot plasma in the post-shock region cools by emitting thermal bremsstrahlung 
X-ray radiation and cyclotron radiation which is observed outside the typical
X-ray range.
Some of the hard X-rays are intercepted by the surface of the WD
and are re-radiated as soft X-ray and EUV regions (e.g. Lamb \& Masters 1979).
Therefore, the WD surface and the post-shock regions are mainly responsible for
the soft (E$\le$ 1 keV) and hard (E $\ge$ 1keV) X-ray emissions of 
polars and intermediate polars, respectively.

Many theoretical models were developed to explain the accretion column (Aizu 1973, King \& 
Lasota 1979, Imamura \& Durisen 1983, Cropper 1990, Wu 1994, Woelk \& 
Beuermann 1996, Cropper et al. 1998, Cropper et al. 1999). 
A blobby accretion model was proposed to interpret the phenomena of flickerings, 
low-level correlations, and the soft excesses often observed in the light curves
and spectra of mCVs (Kuijpers \& Pringle 1982, Frank et al. 1988, Hamuery \& King 1988).
A comprehensive review of the radiative process for mCVs was given by Wu (2000).

Observational information pertaining to an inhomogeneous accretion stream or an 
accretion column which contains dense blobs can be found from studying X-ray light curves, 
particularly, a cross-correlation analysis of energy-dependent light curves. 
Such a study was performed for the first time for AM Her in the optical (V band) and 
soft X-ray regions (Szkody et al. 1980, Tuohy et al. 1981).
Later, analysis of the {\it Einstein} observatory data of AM Her and EF Eri
detected no significant X-ray lag at short time scales but found a low level of 
correlation (Stella et al. 1986, Beuermann et al. 1987, Watson et al. 1987).
Ramsay \& Cropper (2002) reported an anti-correlated hard X-ray lag (20-40 s) for 
the asynchronous polar
BY Cam between soft (0.1-0.3 keV) and hard (1-10 keV) X-ray light
curves of the XMM-{\it Newton} data. They interpreted that this delay
may be due to the obscuration of hard X-rays by optically thick material produced
as a result of collisions between dense blobs and the surface of the WD.

In this paper, we report the results of cross-correlation analysis of
{\it RXTE} observations of a proto-type polar AM Her, which show anti-correlated 
intensity variations and an X-ray lag.
                   
\section{Data Reduction and Analysis}
\label{s:Reduction and Analysis}
The {\it RXTE} satellite has three on-board detectors, 
the Proportional Counter Array (PCA, Jahoda et al. 2006),
the High-Energy X-ray Timing Experiment (HEXTE, Rothschild et al. 1995),
and the All--Sky Monitor (ASM, Levine et al. 1996). 
There are 27 observation data sets for AM Her in the {\it RXTE} archives, referred to 
as ObsIds.
We acquired all the PCA data from the HEASARC public archives and used the 
standard 2 format data to obtain background-subtracted light curves in different
energy bands.
Using the HEASOFT software v6.8, we filtered the data by applying all the required
procedures presented by Christian (2000). 

\subsection{Temporal Analysis}

We have extracted soft (2-4 keV) and hard X-ray (9-20 keV) light curves for 
all the ObsIds of AM Her and performed cross-correlation analysis. 
Most of the data sets showed a strong positive correlation at zero lag.
However, in three data sets (Table 1), the cross-correlation functions (CCFs)
showed a weak correlation or an anti-correlation. 
Using the data observed on 1998 August 4-5, Christian (2000) also performed the
cross-correlation analysis between EUV and X-ray and found no significant
correlation between them.

The soft (2-4 keV) and hard X-ray (9-20 keV) light curves of the three data sets,
along with the hardness ratio (with a bin size of 32 s), are plotted in Figure 1 
in the top-left panel (ObsId 30007-01-01-00), in Figure 1 in the bottom-left panel
(ObsId 20010-01-01-030), and in Figure 2 in the top panel (ObsId 20010-01-01-00).
To study the X-ray lags, we cross-correlated the soft and hard X-ray light curves
using the XRONOS program {\it crosscor} (for details see, e.g., Sriram et al. 2007; Lei et al. 2008).
 This program calculates the error bars by the usual method for 
direct cross-correlation (crosscorr with fast=no option\footnote{http://heasarc.gsfc.nasa.gov/docs/xanadu/xronos/help/crosscor.html}), 
i.e., theoretical error bars for a normal  distribution of errors in the data and error propagation methods are used for larger numbers of intervals.
In most sections, the CCFs showed a positive peak
at a lag of zero. For example, the first section of the light curve for ObsId
30007-01-01-00 displays a strong correlation between the soft and hard X-ray light
curves, and the CCF in the top-right panel of Figure 1 represents a peak at zero 
lag.
Although most sections displayed such strong correlations at zero lag, we found
a weak anti-correlated intensity variation in two sections, the first in 
ObsId 30007-01-01-00 and the second in 20010-01-01-030 (marked by vertical lines).
In ObsId 30007-01-01-00, the second CCF shows an anti-correlation with a lag but 
the CCF is noisy and not very significant, whereas the CCF of 
20010-01-01-030 shows a low level of correlation (see the bottom-right panels in 
Figure 1).

However, in one section of ObsId 20010-01-01-00 (marked by vertical lines in Figure 2),
we found an anti-correlated intensity variation between the soft and hard X-ray light
curves (the other sections of the light curve show a strong correlated intensity
variation). We performed the cross-correlation analysis with 32 s bin data for all the sections of
ObsId 20010-01-01-00 and found that the CCFs show a positive peak at zero lag similar
to the top-right panel of Figure 1, except for the selected section marked by vertical lines.
The marked section was selected on the basis of an opposite intensity variation.
That is, in this section, the soft X-ray intensity varies from 24 c/s to 21 c/s whereas
the corresponding hard X-ray varies from 22 c/s to 44 c/s.

The CCF for the selected section is shown in the middle panel of Figure 2, 
 which indicates an anti-correlation with a lag. We also performed the cross-correlation
analysis for this section with higher time bin (64 s and 96 s) data and found that the lag 
is still present.
 To calculate the X-ray lag time, we fitted an inverted Gaussian function using 
the chi-square minimization method to the CCF of a correlation coefficient (cc) of -0.42$\pm$0.09.
For this fit, we took a negative value for the normalization (-0.42) of the Gaussian
function and fitted the CCF profile spanning -500 s to 500 s (dof = 30).
We found that the hard X-rays are delayed compared to the soft X-rays at a time of 192$\pm$33 s
 (where the measured lag time, i.e. the Gaussian fit offset from zero lag time, is significant 
at a confidence level of 5.8$\sigma$
and the CCF amplitude is about 4$\sigma$. The error is estimated using the criterion $\Delta \chi^{2}$=2.71).

For the section of interest, we also extracted light curves in five different 
energy bands, viz. 4-5 keV, 5-6.5 keV, 6.5-9.0 keV, 9-12 keV and 12-30 keV.
These light curves were cross-correlated with the 2-4 keV light curve.
It was subsequently found that apart from  the 4-5 keV band, all other CCFs showed 
a similar lag at $\sim200$ s.  To study further, we divided this section into two segments
(each segment has of $\sim$1500 sec) and performed the analysis.
The CCFs showed an anti-correlation (the first segment cc = -0.51$\pm$0.1 
and the second segment cc = -0.41$\pm$0.09) with a lag of $\sim$200$\pm$30 s.
Hence, we concluded that there is no energy dependency for the observed lag time and the lag 
is persistent in this section.

\subsection{Spectral Analysis}

To investigate whether there is a significant spectral change during the anti-correlated
intensity variation, we divided the section of interest into two subsections and extracted 
energy spectra (2.5-30.0 keV) from  subsections A and B (top panel of Figure 2).
It was found that the spectra prominently change below and above the energy of 
$\sim$4 keV (the bottom panel of Figure 2). 
Similar kind of study was performed to the other sections (i.e. we divided each section into 
two subsections and checked their spectra by over-plotting the spectra of two subsections)
and found that the other sections do not show such a significant spectral change between the
subsections. There was a change mainly in normalization between the spectra of the given 
subsections.
This study strongly supports that the observed spectral change for the selected section is 
related with the anti-correlated intensity variation.
A similar spectral change was also observed in neutron star and black hole sources
for which anti-correlated X-ray lags were observed
(Choudhury et al. 2005; Lei et al. 2008; Sriram et al. 2007; Sriram et al. 2009). 
This clearly indicates that an astrophysical change of the emission/absorption
region is required to explain the spectral and the anti-correlated intensity 
variations, as well as the observed hard X-ray delay.

To pin down which parameter is responsible for the spectral variation, we fitted
a model to the spectra using the XSPEC software v12.5 (Arnaud 1996). The
model consists of a thermal bremsstrahlung emission and a Gaussian function for
the emission feature around 6.4 keV. The model also includes an absorption component
parameterized with the hydrogen equivalent column density (N$_{H}$) along the line of
sight. The bottom panel of Figure 2 shows unfolded spectra together with the
fitted model functions. The best-fit spectral parameters are tabulated in Table 2.
It was found that during the anti-correlated variation, the emission temperature
increased significantly from kT$_{TB}\sim$13 keV to kT$_{TB}\sim$20 keV, indicating
that the spectrum evolved toward a relatively harder state.
It was also found that the absorption column density increased during the variation.

\section{Discussion and Summary}
\label{s:summary}

Despite the many attempts to do so in the optical to X-ray energy range,
a statistically significant lag phenomenon has not been found in the polar AM Her
(Szkody et al. 1980, Tuohy et al. 1981, Stella et al. 1986, Watson et al. 1987).
The cross-correlation study between {\it Ginga} hard X-ray and optical data showed that
the correlation between them decreases towards bluer wavelengths (Beardmore \& Osborne 1997).
Christian (2000) observed AM Her simultaneously in EUV and X-ray energy bands and the 
cross-correlation study showed no significant correlation.
Stella et al. (1986) found uncorrelated soft and hard X-ray short-term variations from AM Her.
Beuermann et al. (1987) and Watson et al. (1987) observed a low level of correlation
from EF Eri.
A significant hard X-ray lag with a time scale of 20-40 s has thus far only been
reported for the asynchronous polar BY Cam (Ramsay \& Cropper 2002).

In this study, we observed three anti-correlated intensity variations
in total from  specific sections of the light curves of AM Her.
For example, in the section of ObsId 30007-01-01-00, the hard (9-20 keV)
X-ray intensity decreases but the soft (2-4 keV) X-ray intensity increases
(Figure 1). On the other hand, in the two sections of ObsId 20010-01-01-030 and
ObsId 20010-01-01-00, the hard X-ray intensity increases whereas the soft X-ray
intensity decreases (Figure 1 and Figure 2). Such kind of anti-correlated intensity 
variation can be seen in Christian (2000; see Fig. 2 in the paper particularly the
X-ray and EUV light curves at phases 0.2--0.3, 1.0--1.2 \& 1.6--1.7) and in
Beardmore \& Osborne (1997; see their optical and X-ray light curves in Fig. 8).
These anti-correlated intensity variations strongly suggest that our observed variations 
are genuine and on occasion the degree of correlation decreases.
Furthermore, we detected a hard lag of $\sim$190 s from  one of the specific
sections (see \S 2.1), which is statistically significant at
a confidence level of 5.8$\sigma$. 
The hard lag indicates that hard X-ray photons are relatively delayed in their
arrivals at the observer compared with the arrivals of soft X-ray photons.
We also found that both the X-ray emission temperature and the absorption column
density increase during the anti-correlated intensity variation (see \S 2.2).

One may raise a question if these results may be due to a sudden intensity change.
In general, a sudden intensity change in the light curve may result in a spurious
X-ray lag. To test for this possibility, we performed a cross-correlation
analysis after removing the spike-like feature at $\sim 1.6\times10^{4}$ s
(between subsection A and B in Figure 2), finding that the CCF profile together
with the corresponding hard X-ray lag ($\sim$190 s) remain.
This signifies that the obtained hard X-ray lag is not due to the spike-like feature. 
%It may happen that during the observed lag, the background spectrum varied and is not properly 
%modeled and estimated. Hence we carefully checked the background spectra of all the sections 
%using PCA and HEXTE data for ObsId 20010-01-01-00.
%We found that the background spectra showed no changes compared to the background
%spectrum  of the selected section (for which the hard X-ray lag was observed). 
%This unarguably indicates that the detected lag feature is associated with the source variability.

The observed hard X-ray lag can be explained in principle if a dense material
exists that obscures only the region that emits the hard X-ray for a short duration. 
Therefore, we can consider dense blobs as possibly involved intermittently in 
an accretion stream or flow.
Such a scenario, however, is unlikely because it requires a very special
geometrical condition to obscure only the hard X-ray emitting region at the
magnetospheric boundary of the WD, i.e., such a specific configuration is
difficult to attain in this system for a short duration.
Furthermore, with this scenario, it is difficult to interpret the spectral
change as well as the anti-correlated intensity variation obtained in \S 2.

The {\it RXTE} PCA energy range (2-30 keV) is sensitive to the typical bremsstrahlung 
continuum radiation of the post-shock region but insensitive to the UV/soft X-ray 
component often observed in AM Her with a temperature of 20-40 eV. Therefore, it is
reasonable to consider an accretion column containing dense blobs. 
We conjecture that the blobs can disturb the temperature and density
structures of the post-shock region. 
For example, if some of the material of the blobs splashes back into the accretion
column after colliding with the WD's surface, this material may take on
a configuration that produces the observed hard X-ray lag.
Such a scenario was proposed by Ramsay \& Cropper (2002) in their interpretation
of the hard X-ray lag of the source BY Cam.
Since the hard X-ray lag is not detected in all occasions, this indicates that
the splashed material stays at the hard X-ray emitting region for a short duration.
We speculate that the duration of disturbance in the accretion column would depend on
the inflow time of the blobs.
We also postulate that both the anti-correlated intensity variation (Figure 2) and
the spectral change (bottom panel of Figure 2 and Table 2) can be explained if some of 
the splashed material can be heated up to emit hard X-ray radiation.
It is, however, difficult to explain our results consistently because there is a 
lack of theoretical understanding regarding how the splashed back material disturbs 
the temperature and density structures of the accretion column.

In summary, we report a hard X-ray lag of $192\pm33$ s in a specific section of
the energy-dependent light curve of AM Her which is statistically significant at
a confidence level of 5.8$\sigma$. The corresponding hard (9-20 keV) and
soft (2-4 keV) X-ray intensities display a strong anti-correlated variation.
We also found that both the X-ray emission temperature and the absorption column
density increase during the anti-correlated intensity variation (see \S 2.2).
The hard X-ray lag is detected for the first time in the given energy range,
and it is the longest lag among those reported in polars and intermediate
polars.
In two other observations, the cross-correlation functions of energy-dependent
light curves showed a low level of correlation, which is consistent with the earlier
results performed in a different energy range.
It is proposed that a similar study of the other mCVs is important to determine
the astrophysical conditions of the post-shock region.

%Future observations with {\it Suzaku}, XMM-{\it Newton} along with RXTE would be helpful 
%to decipher the anti-correlated intensity variations and possible lags in different
%energy bands from magnetic cataclysmic variables.

\begin{acknowledgements}
%\acknowledgements 
We are grateful to the anonymous referee for useful comments.
This research has made use of data obtained through the HEASARC Online Service, 
provided by NASA/GSFC in support of the NASA High Energy Astrophysics Programs.
\end{acknowledgements}
%%%%%%%%%%%%%%%%%%%%%%%%%%%%%%%%%%%%%%%%%%%%%%%%%%%%%%%%%%%%%%%%%%%%%%%%%%%%%%%%%
\begin{table}
\begin{minipage}[t]{\columnwidth}
\caption{The three RXTE observations of AM Her used for our cross-correlation study.}
\label{tab1}
\centering
\renewcommand{\footnoterule}{}
\begin{tabular}{cccc}
\hline
ObsId & Date& \multicolumn{2}{c}{ Time (UT)}\\
& &Start& Stop\\
\hline

30007-01-01-00&1998 Aug 4&09:48:12&17:18:14\\
20010-01-01-00&1998 Aug 24&09:47:29&16:41:14\\
20010-01-01-030&1998 Aug 26&22:11:20 &06:00:07\\

\hline
\hline
\end{tabular}
\end{minipage}
\end{table}

\begin{table}
\begin{minipage}[t]{\columnwidth}
\caption{Best-fit spectral parameters for the spectra of the A and B subsections. 
The attached errors are at a 90\% confidence level.} 
\label{tab1}
\centering
\renewcommand{\footnoterule}{}
\begin{tabular}{ccc}
\hline
Parameters&\multicolumn{2}{c}{ObsId 20010-01-01-00} \\
\hline
&A&B\\
\hline
\hline
N$_{H}$($\times$ 10$^{22}$ cm$^{-2}$)\footnote{Hydrogen equivalent column density} &$1.4\pm0.3$ & 3.2$\pm$0.5\\

$kT_{TB}$ (keV)\footnote{Plasma temperature of thermal bremsstrahlung model}&$13.6\pm1.1$ &21.2$\pm$1.5\\
$N_{TB}$\footnote{Normalization of thermal bremsstrahlung model}&0.041$\pm$0.001&0.048$\pm$0.001\\

$E_{line}$ (keV)\footnote{Energy of Gaussian emission line}&6.62$\pm$0.07 & 6.58$\pm$0.04\\
EW (eV)\footnote{Width of the emission line}& 326.1$\pm$0.7 &368.2$\pm$1.6\\

$\chi^{2}$/dof&45.75/53 &43.41/53 \\

\hline
\hline
\end{tabular}
\end{minipage}
\end{table}

%%%%%%%%%%%%%%%%%%%%%%%%%%%%%%%%%%%%%%%%%%%%%%%%%%%%%%%%%%%%%%%%%%%%%%%%%
\begin{figure}
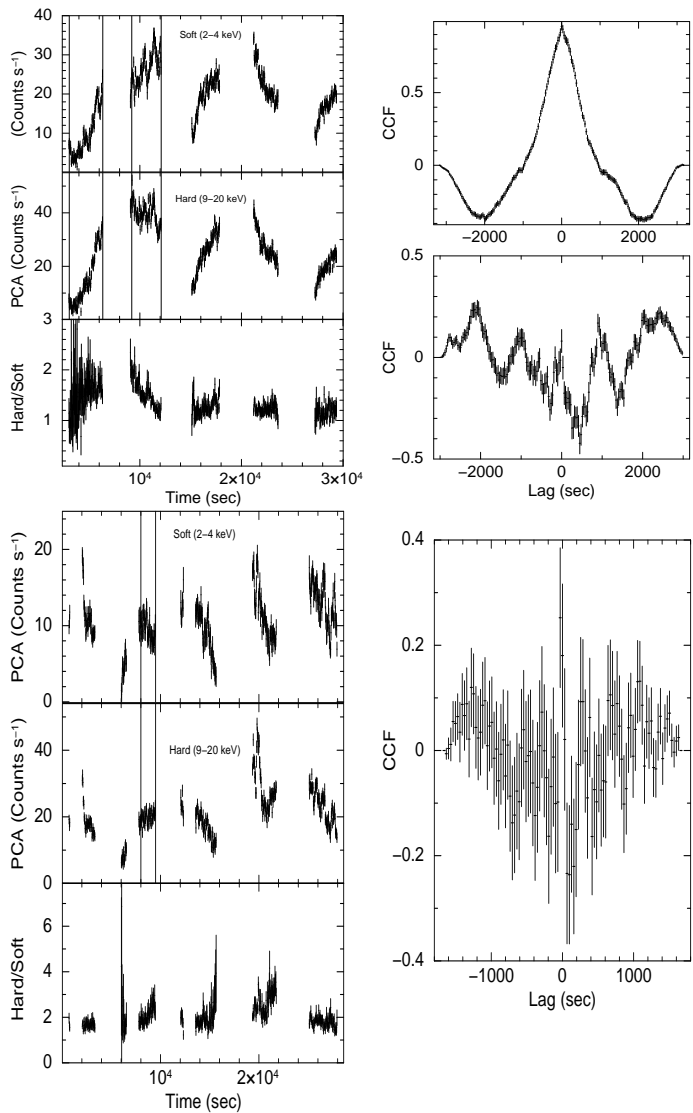

%\centering
\resizebox{\hsize}{!}
%\resizebox{\textwidth}{1}   
%\includegraphics[height=15cm,width=10cm,angle=-90]{fig1a_lc_ccor.ps}
%\includegraphics[height=15cm,width=10cm,angle=-90]{fig1b_lc_ccor.ps}
{\includegraphics[height=8cm,width=6cm,angle=-90]{fig1a_lc_ccor.ps}}\\%for printer format for aa
{\includegraphics[height=9cm,width=8cm,angle=-90]{fig1b_lc_ccor.ps}}%for printer format for aa
%{\includegraphics[height=8cm,width=5cm,angle=-90]{fig1a_lc_ccor.ps}}\\%for Referee format for aa 
%{\includegraphics[height=19cm,width=10cm,angle=-90]{fig1b_lc_ccor.ps}}%for Referee format for aa
\caption{Top: The PCA soft (2-4 keV) and  hard (9-20 keV) X-ray light curves of
AM Her, along with the hardness ratios (ratios of 9-20 keV counts to 2-4 keV counts), 
are shown for ObsId 30007-01-01-00. 
The vertical lines in the light curves represent the sections used in our 
cross-correlation study.
The cross-correlation functions (CCF) for the first and second sections
are displayed in the two top-right panels, respectively.
Bottom: Similar to the top panel but for ObsId 20010-01-01-030.
%The light curves, hardness ratios, and the CCF for the selected section 
%(marked by vertical lines) are shown for ObsId 20010-01-01-030.
} 
       %\label{Fig1}
 \end{figure}

\begin{figure}
%\centering
\resizebox{\hsize}{!}
{\includegraphics[height=8cm,width=6cm,angle=-90]{fig2a_lc.ps}}\\%for printer format for aa
{\includegraphics[height=10cm,width=4cm,angle=-90]{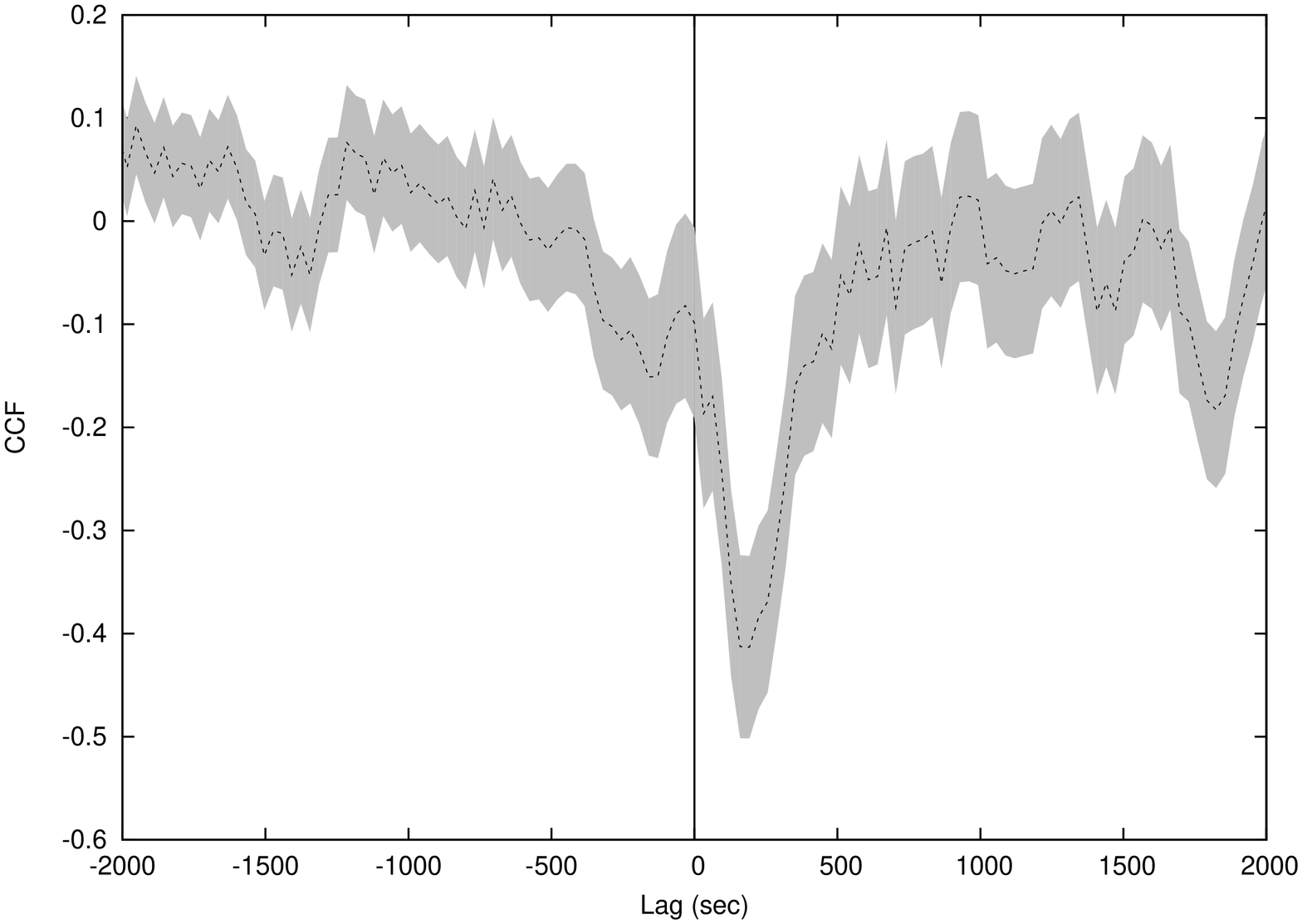}}\\%for printer format for aa
{\includegraphics[height=4cm,width=10cm,angle=0]{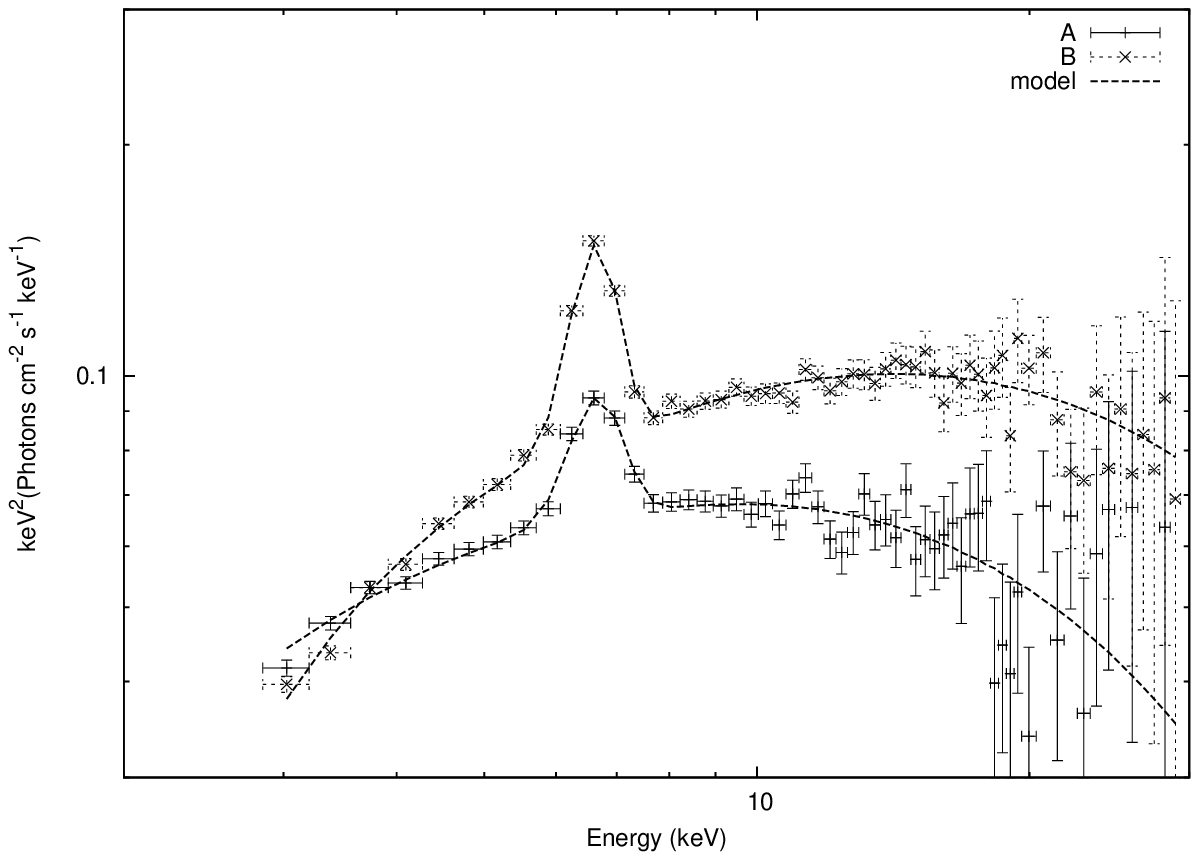}}%for printer format for aa
%{\includegraphics[height=8cm,width=5cm,angle=-90]{fig2a_lc.ps}}\\%for Referee  format for aa
%{\includegraphics[height=19cm,width=6cm,angle=-90]{fig2b_ccor.ps}}\\%for Ref format for aa
%{\includegraphics[height=6cm,width=19cm,angle=0]{fig2c_specAB_fit.ps}}%for Ref format for aa
     \caption{Top: The PCA soft (2-4 keV) and hard (9-20 keV) X-ray light
curves of AM Her, along with the hardness ratios, are shown for ObsId 20010-01-01-00.
The solid vertical lines represent the section used in our cross-correlation study.
We extracted the energy spectra from the subsections A and B (see text). 
Middle: The cross-correlation function (CCF) for the selected section is presented. 
The dashed line represents the mean of the CCF and the gray shaded region 
corresponds to the error range obtained from the {\it crosscor} (see text). 
%(with option fast=no) utility, 
%{\bf which calculates the error bars by the usual method for 
%direct cross-correlation (crosscorr with fast=no option\footnote{see http://heasarc.gsfc.nasa.gov/docs/xanadu/xronos/help/crosscor.html}
%i.e., theoretical error bars for a normal  distribution of errors in the data. Error propagation methods are used for larger numbers of intervals}. 
The vertical line represents a lag of zero.
Bottom:The unfolded spectra (for the section A \& B, see top panel) are shown together with the fitted model functions.} 
       \end{figure}

%\begin{figure}
%\centering
%\resizebox{\hsize}{!}
%\includegraphics[height=18cm,width=10cm,angle=-90]{fig3a_piv.ps}
%\includegraphics[height=10cm,width=18cm,angle=0]{fig3b_specAB_fit.ps}
%{\includegraphics[height=5cm,width=5cm,angle=-90]{fig3a_piv.ps}}
%{\includegraphics[height=10cm,width=15cm,angle=0]{fig3b_specAB_fit.ps}}
%\caption{%Top: The spectra extracted from  subsections A (marked as circles) and B 
%(marked as stars) are plotted.  
%Bottom: 
%The unfolded spectra are shown together with the fitted model functions. 
%} 
%\label{Fig3}
 %\end{figure}
%%%%%%%%%%%%%%%%%%%%%%%%%%%%%%%%%%%%%%%%%%%%%%%%%%%%%%%%%%%%%%%%%%%%%%%%%%%%%%%%%%%%%%%%

%%%%%%%%%%%%%%%%%%%%%%%%%%%%

\begin{thebibliography}{}
\small
\bibitem[Aizu 1973]{a73}Aizu, K. 1973, Prog. Theor. Phys., 49, 184
\bibitem[Arnaud 1996]{ar96}Arnaud, K. A. 1996, ADASS, 5, 17
\bibitem[Beardmore \& Osborne]{1997}Beardmore, A. P., \& Osborne, J. P. 1997, MNRAS, 290, 145
\bibitem[Beuermann et al. 1987]{B87}Beuermann, K., Stella, L., \& Patterson, J. 1987, \apj, 316, 360
\bibitem[Christian 2000]{C00}Christian, D. J. 2000, \aj, 119, 1930
\bibitem[Choudhury et al. 2005]{choudhury05} Choudhury, M., et al. 2005, \apj, 631, 1072

\bibitem[Cropper 1990]{cr90}Cropper, M. 1990, {\it Space Sci. Rev.}, 54, 195 
\bibitem[Cropper et al. 1998]{cr98}Cropper, M., Ramsay, G., \& Wu, K. 1998, \mnras, 293, 222
\bibitem[Cropper et al. 1999]{cr99}Cropper, M., et al. 1999, \mnras, 306, 684
\bibitem[Frank et al. 1988]{F88}Frank, J., King, A. R., \& Lasota, J. P. 1988, A\&A, 193, 113
\bibitem[Hameury \& King 1988]{HK98}Hameury, J. M., \& King, A. R. 1988, MNRAS, 1988, 235, 433
\bibitem[Imamura \& Durisen]{ID83} Imamura, J. N., \& Durisen, R. H. 1983, \apj, 268, 291

\bibitem[Jahoda et al. 2006]{jahoda} Jahoda, K., et al. 2006, \apjs, 163, 401
\bibitem[King \& Lasota 1979]{KL79}King, A. R., \& Lasota, J. P. 1979, \mnras, 188, 653 
\bibitem[Kuijpers \& Pringle]{KP82}Kuijpers, J., \& Pringle, J. E. 1982, A\&A, 114, L4
\bibitem[Lamb \& Master]{lm79}Lamb, D. Q., \& Master, A. R. 1979, \apj, 234, L117
\bibitem[Lei et al. 2008]{lei08}Lei, Y. J., et al. 2008, \apj, 677, 461
\bibitem[Levine et al. 1996]{L96}Levine, A. M., et al. 1996, \apj, 469, L33

\bibitem[Rothschild et al. 1995]{rothschild} Rothschild, R. E., et al. 1995, \procspie, 2518, 13
\bibitem[Ramsay \& Cropper]{RC02}Ramsay, G., \& Cropper, M. 2002, \mnras, 334, 805
\bibitem[Sriram et al. 2007]{sriram07} Sriram, K., et al. 2007, \apj, 661, 1055
\bibitem[Sriram et al. 2009]{sriram09} Sriram, K., Agrawal, V. K., \& Rao, A. R. 2009, RAA, 9, 901
\bibitem[Stella et al. 1986]{s86} Stella, L., Beuermann, K., \& Patterson, J. 1986, \apj, 306, 225
\bibitem[Szkody et al. 1980]{sz80}Szkody, P., et al. 1980, \apj, 241, 1070
\bibitem[Tuohy et al. 1981]{tu81}Tuohy, I. R., Mason, K. O., Garmire, G. P. \& Lamb, F. K. 1981, \apj, 245, 183
\bibitem[Watson et al. 1987]{w87}Watson, M. G., King, A. R., \& Williams, G. A. 1987, \mnras, 226, 867
\bibitem[Woelk \& Beuermann 1996]{Wo96}Woelk, U., \& Beuermann, K. 1996, A\&A, 306, 232
\bibitem[Wu 1994]{wu01}Wu, K. 1994, Proc. Astron. Soc. Aust., 11, 61
\bibitem[Wu 2000]{wu0}Wu, K. 2000, {\it Space Sci. Rev}, 93, 611

\end{thebibliography}
\end{document}